\documentclass[%
 aip,
rsi,%
 amsmath,amssymb,
 reprint,%
]{revtex4-1}

\usepackage{graphicx}
\usepackage{dcolumn}
\usepackage{bm}
\usepackage{float}
\usepackage{array}
\usepackage{textcomp}

\begin{document}
\newcolumntype{C}[1]{>{\centering\arraybackslash}p{#1}}
\newcommand{\beginsupplement}{%
        \setcounter{table}{0}
        \renewcommand{\thetable}{S\arabic{table}}%
        \setcounter{figure}{0}
        \renewcommand{\thefigure}{S\arabic{figure}}%
     }
     
\preprint{AIP/123-QED}

\title[Dielectric loss extraction for superconducting microwave resonators]{Dielectric loss extraction for superconducting microwave resonators}

\author{C.R.H. McRae}
\email{coreyrae.mcrae@colorado.edu}
\affiliation{
Department of Physics, University of Colorado, Boulder, Colorado 80309, USA}
\affiliation{ 
National Institute of Standards and Technology, Boulder, Colorado 80305, USA
}%
\author{R.E. Lake}
\affiliation{ 
National Institute of Standards and Technology, Boulder, Colorado 80305, USA
}%
\author{J.L. Long}%
\affiliation{
Department of Physics, University of Colorado, Boulder, Colorado 80309, USA}
\affiliation{ 
National Institute of Standards and Technology, Boulder, Colorado 80305, USA
}%
\author{M. Bal}%
\affiliation{
Department of Physics, University of Colorado, Boulder, Colorado 80309, USA}
\affiliation{ 
National Institute of Standards and Technology, Boulder, Colorado 80305, USA
}%
\author{X. Wu}%
\affiliation{Lawrence Livermore National Laboratory, Livermore, California 94550, USA
}%
\author{B. Jugdersuren}%
\affiliation{Jacobs Engineering Group, Hanover, MD 21076 USA
}%
\author{T.H. Metcalf}%
\affiliation{Naval Research Laboratory, Washington, DC 20375 USA
}%
\author{X. Liu}%
\affiliation{Naval Research Laboratory, Washington, DC 20375 USA
}%
\author{D.P. Pappas}%
\affiliation{ 
National Institute of Standards and Technology, Boulder, Colorado 80305, USA
}%

\date{\today}

\begin{abstract}
The investigation of two-level-state (TLS) loss in dielectric materials and interfaces remains at the forefront of materials research in superconducting quantum circuits. We demonstrate a method of TLS loss extraction of a thin film dielectric by measuring a lumped element resonator fabricated from a superconductor-dielectric-superconductor trilayer.  We extract the dielectric loss by formulating a circuit model for a lumped element resonator with TLS loss and then fitting to this model using measurements from a set of three resonator designs: a coplanar waveguide resonator, a lumped element resonator with an interdigitated capacitor, and a lumped element resonator with a parallel plate capacitor that includes the dielectric thin film of interest. Unlike the commonly used single measurement technique, this method allows accurate measurement of materials with TLS loss lower than $10^{-6}$. We demonstrate this method by extracting a TLS loss of $1.00 \times 10^{-3}$ for sputtered $\mathrm{Al_2O_3}$ using a set of samples fabricated from an $\mathrm{Al/Al_2O_3/Al}$ trilayer. We compare this method to the single measurement technique and observe a difference of 11$\%$ in extracted loss of the trilayer.
%
\end{abstract}

\maketitle

Two-level-state (TLS) loss is the dominant form of loss at millikelvin temperatures and single photon powers in superconducting quantum circuits.~\cite{Martinis2005} TLS loss is a type of dielectric loss that occurs due to an interaction with an electric field, and is generated in bulk dielectrics and interfaces between materials in superconducting quantum circuits.~\cite{Gao2008,Calusine2018} Materials improvements in superconducting quantum computing have largely focused on reducing the density and total loss of TLS by improving fabrication,~\cite{Sandberg2012,Quintana2014,Bruno2015} identifying high- and low-loss regions~\cite{Wisbey2010,Calusine2018,Vissers2010} and modifying circuit design to reduce participation of lossy materials.~\cite{Chang2013,Wang2015}

The total loss in a superconducting microwave resonator can be written as:
\begin{equation}
\label{eqn:totalloss}
     \tan{\delta}= { 1 \over Q_{i} } = F \tan{\delta_{\mathrm{TLS}}} + { 1 \over Q_{\mathrm{HP}} }
\end{equation}
where $Q_{i}$ is the internal quality factor of the resonator and is equal to the inverse of the total loss in the resonator $\tan{\delta}$, $F \tan{\delta_{\mathrm{TLS}}}$ is the TLS loss with $F$ denoting the filling factor of the TLS material, and $1 \over Q_{\mathrm{HP}}$ is the high power loss. High power loss is generally small and power-independent in the operational regime of a superconducting quantum circuit, whereas TLS loss has a distinctive power dependence as well as a temperature dependence. It is also of interest to note that high power loss is related to the inductance of the circuit, whereas TLS loss is related to its capacitance.

Much is still uncertain about the origins and behavior of TLS.~\cite{Muller2017} The general model for weak-field TLS loss as a function of power and temperature is:~\cite{Earnest2018,Burnett2018,Richardson2016}
\begin{equation}
\label{eqn:genTLSloss}
    F \tan{\delta_{\mathrm{TLS}}} = F \tan{\delta_{\mathrm{TLS}}^0} {\tanh({\hbar \omega_0 \over 2 k_B T}) \over (1 + ({\langle n \rangle \over n_c}))^\beta}.
\end{equation}
where $F \tan{\delta_{\mathrm{TLS}}^0}$ is the TLS loss of the system at zero power and temperature ($\langle n \rangle = 0$ and $T = 0$), $\omega_0$ is the angular resonance frequency, and $\beta$ is a variable determined by TLS population densities, but is usually close to 0.5. TLS become saturated at high powers, and therefore do not contribute to high power loss. As power decreases in the circuit, TLS loss participation increases until it flattens around single photon powers near the critical photon number $n_{c}$. $\tan{\delta_{\mathrm{TLS}}^0}$ can be seen as an intrinsic value of the TLS material in question, and varies with properties of the material such as deposition parameters, surface treatments, and crystallinity.~\cite{Wang2009,Sage2011,Vissers2010,Megrant2012}

Only capacitive components contribute to TLS loss.~\cite{Vissers2012} In the past, dielectric loss has been measured using coplanar waveguide (CPW) resonators,\cite{OConnell2008,Sage2011,McRae2018} lumped element (LE) resonators with parallel plate capacitors (PPCs),\cite{Cho2013,Deng2014} and LE resonators with interdigitated capacitors (IDCs).~\cite{Vissers2012} In one strategy, the filling factor of the material is determined through simulation.~\cite{OConnell2008,Calusine2018}

It has been previously assumed that, in a lumped element resonator with a PPC, a negligible amount of capacitance comes from the inductor,~\cite{OConnell2008,Weber2011} so that the total TLS loss of the resonator is roughly equal to the TLS loss of the PPC. Using this assumption, a single resonator design can be measured to determine the TLS loss of a dielectric material in the PPC. This ``single measurement technique" is valid when the participation and/or loss of the material in the capacitor dominates the loss of other components in the resonator.

The identification of low loss dielectrics ($\tan{\delta^0_{\mathrm{TLS}}} \lesssim 10^{-6}$) for use as substrates, junction insulators, and spacer materials for three-dimensional integration would allow for the expansion of possible circuit architectures. The implementation of a low loss dielectric could drastically decrease the qubit footprint from one millimeter to micrometers. In this work, we demonstrate that the single measurement technique is not sensitive enough to determine the loss of low loss materials, and a method to remove losses from other circuit components is necessary.

We present a technique to extract the TLS loss of a given thin film dielectric material using measurements of three resonator designs: an LE resonator with a PPC, an LE resonator with an IDC, and a CPW resonator. We apply this technique to measure the TLS loss of sputtered $\mathrm{Al/Al_{2}O_{3}/Al}$ trilayers in order to report a TLS loss value of $1.00 \times 10^{-3}$ with a difference of 11$\%$ from the single measurement technique. We also outline the design and materials regimes where assumptions of the single measurement technique no longer apply and the losses of other resonator components must be addressed.

\begin{figure}
\includegraphics[width=85mm]{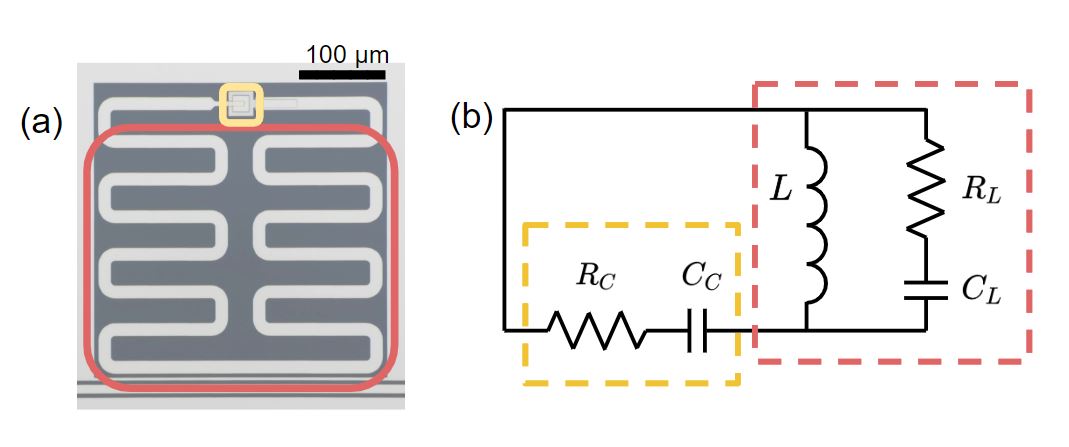}
\caption{\label{fig:circuit} (a) Optical micrograph of $\mathrm{Al/Al_{2}O_{3}/Al}$ PPC LE resonator. (b) RLC circuit representing TLS loss in a superconducting lumped element resonator. Yellow (light grey) rectangle denotes the LE capacitor, and red (dark grey) rectangle denotes LE inductor.}
\end{figure}

\begin{figure*}
\includegraphics[width=150mm]{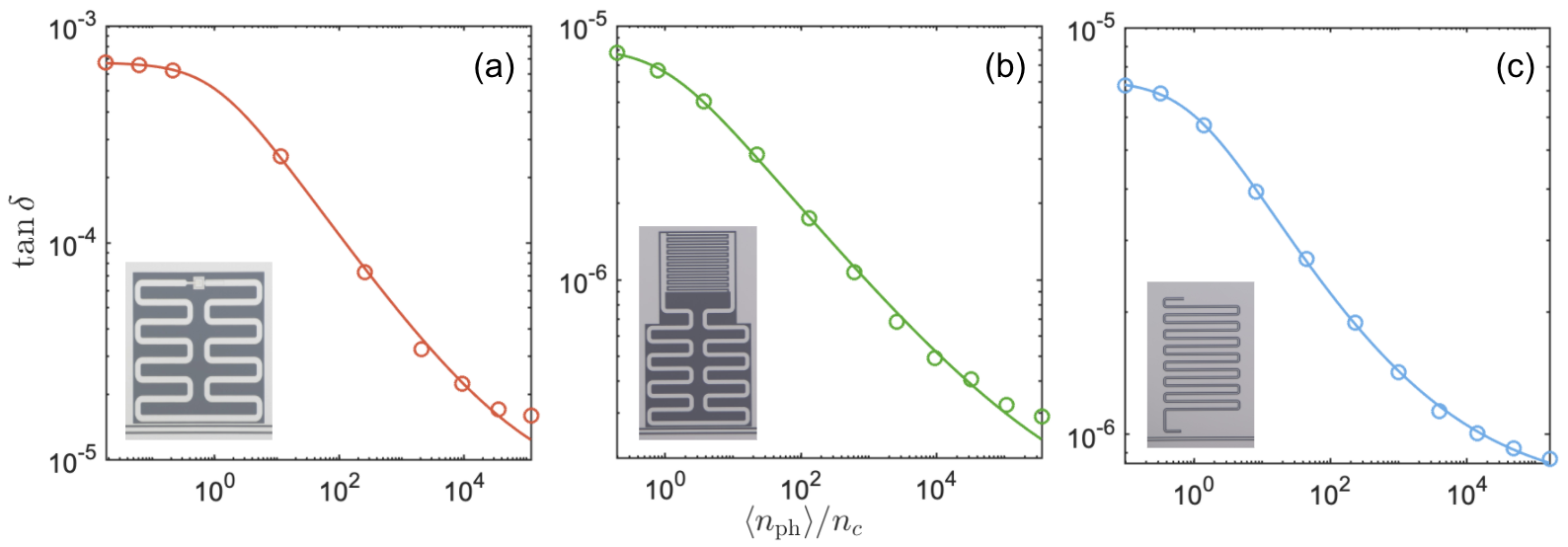}
\caption{\label{fig:measAlOx}\label{fig:measIDCCPW}Photon number sweeps for (a) an LE PPC resonator (device A), (b) an LE IDC (device B) and (c) a CPW (device C) resonator. Data (circles) of loss $\tan{\delta}$ as a function of fractional mean photon number $\langle n \rangle / n_c$, and fit to Eqn.~\ref{eqn:totalloss} (solid lines) is shown. For space reasons, optical micrographs of similar devices are shown as insets. Shown here: an LE PPC resonator of $N$ = 7 (rather than $N$ = 17), an LE IDC resonator of $N$ = 7 (rather than $N$ = 13), and a compressed CPW resonator, where $N$ is number of inductor arm pairs in design.}
\end{figure*}

The material under test is a sputtered $\mathrm{Al/Al_{2}O_{3}/Al}$ trilayer deposited at the Naval Research Laboratory. 50 nm of Al, 50 nm of $\mathrm{Al_{2}O_{3}}$, and 50 nm of Al were deposited consecutively at room temperature without breaking vacuum, with a base pressure of $6 \times 10^{-6}$ Pa. The $\mathrm{Al/Al_{2}O_{3}/Al}$ trilayer is patterned into a PPC and incorporated into an LE resonator (Fig.~\ref{fig:circuit}~(a)) in order to perform TLS loss measurements. An LE IDC resonator (Fig.~\ref{fig:measIDCCPW}~(b) inset) and a CPW resonator (Fig.~\ref{fig:measIDCCPW}~(c) inset) are also measured in this work. These resonators are fabricated on the same wafer as the LE PPC resonators and are defined with liftoff of electron-beam-evaporated Al in the same step as the inductors in the LE PPC resonators. More details on fabrication and geometry can be found in Table~\ref{tab:meas} and in the supplementary material.

TLS loss in a superconducting lumped element resonator can be modeled by an RLC circuit. When considering TLS loss exclusively, only capacitive components have associated resistive components. Each lossy capacitor is modeled as a lossless ideal capacitor with equivalent series resistance (ESR) representing the TLS loss of that component. In this way, the lumped element capacitor is represented by an ideal capacitor of capacitance $C_{C}$ with an associated ESR of resistance $R_{C}$.

The inductor in a non-ideal resonator is not a purely inductive component. Some amount of stray capacitance will always be present within the inductor itself or to ground. Therefore, the inductor can be modeled as a pure lossless inductor $L$ with a capacitor of capacitance $C_{L}$ and ESR of resistance $R_{L}$. A diagram of the full circuit is shown in Fig.~\ref{fig:circuit}~(b). 

We can write the total capacitance of the resonator as $C_{\mathrm{tot}} = C_{C} + C_{L}$. Then, the total TLS loss of the resonator at zero power and temperature can be written as:
\begin{equation}
F_{\mathrm{tot}}\tan{\delta_{\mathrm{tot}}} = {C_{C} \over C_{\mathrm{tot}}} F_{C} \tan{\delta_{C}} + {C_{L} \over C_{\mathrm{tot}}} F_{L} \tan{\delta_{L}}
\end{equation}
where $F_{\mathrm{tot}}$, $F_{C}$, and $F_{L}$ are filling factors of the TLS material, and ${C_{C} \over C_{\mathrm{tot}}}$ and ${C_{L} \over C_{\mathrm{tot}}}$ are the participation ratios of the capacitor and inductor respectively, which is equivalent to the fraction of the total resonator capacitance in each element. Here we are omitting the ``0" superscript for brevity, but $F_{\mathrm{tot}}\tan{\delta_{\mathrm{tot}}} = F\tan{\delta_{\mathrm{\mathrm{TLS}}}^0}$ as in Eqn.~\ref{eqn:genTLSloss}.

In order to determine the amount of loss associated with the LE inductor and capacitor respectively, $C_{C}$ and $C_{L}$ must be known. These can be determined through a combination of simulation and measurement as demonstrated in this work. By performing measurements of strategically designed devices, the loss of a single component within the resonator can be determined.

\begin{table*}
\caption{\label{tab:meas}Parameters for three measured devices. $F \tan{\delta_{\mathrm{TLS}}^{0}}$: measured TLS loss. $f_{0}$: measured resonance frequency. $N$: number of inductor arm pairs in design. $g_{c}$: designed coupling gap. $C_{C}$: capacitance of capacitor extracted from measurement, simulation, and analytical methods. $C_L$: capacitance of inductor extracted from a combination of measurement and simulation. $L$: inductance of inductor determined by simulation.}
\begin{ruledtabular}
\begin{tabular}{cccccccccc}
Design & Material & Label & $F \tan{\delta_{\mathrm{TLS}}^{0}} (\times 10^{-6})$ & $f_{0}$ (GHz) & $N$ & $g_{c}$ (\textmu m) & $C_{C}$ (fF) & $C_L$ (fF) & $L$ (nH) \\
\hline
LE PPC & $\mathrm{Al/Al_2O_3/Al}$ & A & 920 $\pm$ 7 & 3.7464 & 17 & 3 & 727.7 & 82.2 & 2.42\\
LE IDC & Planar Al & B & 8.9 $\pm$ 0.1 & 6.3798 & 13 & 30 & 34.7 & 64.4 & 1.87 \\
CPW & Planar Al & C & 8.42 $\pm$ 0.06 & 4.5548 & - & - & - & - & - \\
\end{tabular}
\end{ruledtabular}
\end{table*}

The loss of the PPC can be determined from a set of three devices: an LE resonator with a PPC, an LE resonator with an IDC, and a CPW resonator. The PPC LE resonator loss is composed of inductor and PPC loss, as: 
\begin{equation}
\label{eqn:PPCloss}
F_{A}\tan{\delta_{A}} = {C_{\mathrm{TLS}} \over C_{A}} F_{\mathrm{PPC}} \tan{\delta_{\mathrm{PPC}}} + {C_{L} \over C_{A}} F_{L} \tan{\delta_{L}}
\end{equation}
where the first term is PPC loss and the second term is inductor loss. We refer to the PPC LE resonator as device A. The single measurement technique requires the assumption that:
\begin{equation}
{C_{\mathrm{TLS}} \over C_{A}} F_{\mathrm{PPC}} \tan{\delta_{\mathrm{PPC}}} \gg {C_{L} \over C_{A}} F_{L} \tan{\delta_{L}}
\end{equation}
and thus:
\begin{equation}
F_{A}\tan{\delta_{A}} \sim {C_{\mathrm{TLS}} \over C_{A}} F_{\mathrm{PPC}} \tan{\delta_{\mathrm{PPC}}}.
\end{equation}
Under this assumption, only a measurement of device A is needed in order to determine the loss of the PPC. However, if this assumption does not hold, measurements of related devices are necessary in order to extract the PPC loss. This ``dielectric loss extraction method" is outlined below.

We can measure a LE IDC resonator (device B) with the same inductor as above. Then we see:
\begin{equation}
\label{eqn:IDCloss}
F_{B}\tan{\delta_{B}} =  {C_{\mathrm{IDC}} \over C_{B}} F_{\mathrm{IDC}}\tan{\delta_{\mathrm{IDC}}} + {C_{L} \over C_{B}} F_{L} \tan{\delta_{L}}.
\end{equation}

We can use these measurements to solve for the PPC loss if we also know ${C_{\mathrm{IDC}} \over C_{B}} F_{\mathrm{IDC}}\tan{\delta_{\mathrm{IDC}}}$. An estimation of this term can be made by measuring a CPW resonator that mimics the TLS loss environment of the IDC by having the same CPW gap and width as the fingers of the IDC. Then:
\begin{equation}
\label{eqn:CPWloss}
     F_{\mathrm{CPW}}\tan{\delta_{\mathrm{CPW}}} \sim F_{\mathrm{IDC}}\tan{\delta_{\mathrm{IDC}}}.
\end{equation}

If the capacitances of each element are known, then from these three equations, and using the fact that $F_{\mathrm{PPC}} = 1$, we can solve for $\tan{\delta_{\mathrm{TLS}}}$. An application of this method is shown below, where the loss of an $\mathrm{Al_2O_3}$ PPC is extracted by measuring, simulating and modeling PPC, IDC, and CPW structures.

\begin{figure}
\includegraphics[width=85mm]{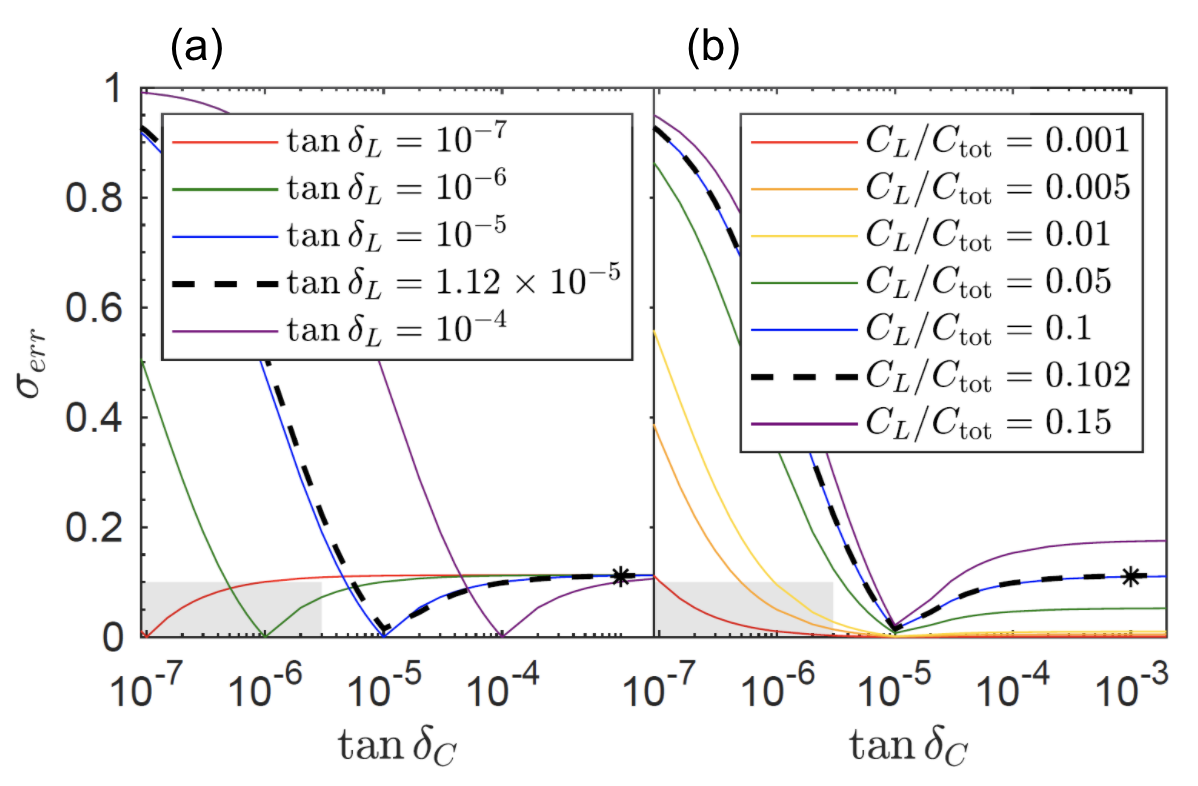}
\caption{\label{fig:error}Error using the single measurement technique $\sigma_{\mathrm{err}}$ as a function of capacitor loss $\tan{\delta_C}$. Black star represents parameter set associated with the measurements in this work. Dashed black line represents the measurements from this work, with $F_{C}\tan{\delta_C}$ left as a free parameter. Grey boxes denote the regime where accurate measurements ($\sigma_{err} \lesssim 0.1$) of low loss dielectrics ($\tan{\delta^0_{\mathrm{TLS}}} \lesssim 10^{-6}$) are possible. For simplicity, $F_{C}$ and $F_{L}$ are assumed to be 1. Comparison of (a) inductor loss $F_{L}\tan\delta_{L} = \tan\delta_{L}$, with $C_L/C_{\mathrm{tot}} = 0.102$, and (b) inductor participation ratio, with $F_{L}\tan\delta_{L} = \tan\delta_{L} = 1.12 \times 10^{-5}$.}
\end{figure}

The inductor design is simulated in Sonnet (see the supplementary material for details) with a varying LE capacitance $C_C$ in order to extract the resonance frequency $f_0$. The frequency response is given by:
\begin{equation}
\label{eqn:Ccextraction}
f_{0} = {1 \over 2 \pi \sqrt{L (C_{C} + C_{L})}}.
\end{equation}
This equation is used to extract the inductance $L$ and capacitance $C_L$ of the inductor. It is possible to engineer the inductor to minimize $C_L$ and maximize the participation of the capacitor, thus increasing the accuracy of the single measurement technique (see the supplementary material for examples). Simulated values for measured resonators in this experiment are given in Table~\ref{tab:meas}.

The capacitances of the experimental $\mathrm{Al_{2}O_{3}}$ PPC and planar IDC are determined by taking the measured resonance frequencies of a series of resonators of each type and solving for the capacitance of the capacitor, $C_{C}$ in the model above, where $L = L_{\mathrm{offset}} + L_{\mathrm{arm}} N$, and $C_L = C_{L,\mathrm{offset}} + C_{L,\mathrm{arm}} N$. $L$ and $C_L$ are determined by Sonnet simulations of an LE resonator with varying $C_C$ and number of inductor arm pairs $N$. $C_C$ is then determined by comparing measured resonance frequencies to Eqn.~\ref{eqn:Ccextraction}. Note that we find a residual $N$-dependent component when performing this comparison, which acts as a correction term within $C_{L,\mathrm{arm}}$. We attribute this to a slight difference between the simulated and fabricated inductor design; the simulated inductor arms have square corners in order to reduce simulation complexity, whereas the fabricated inductor arms have rounded corners in order to prevent current crowding.

Using this method with simulated values, we obtain the $C_C$ values shown in Table~\ref{tab:meas} for the PPC. We are able to perform the calculation above due to the assumption that the PPC introduces negligible inductance to the circuit. For the IDC, it is more accurate to calculate $C_C$ analytically.~\cite{Bahl2013}

An LE PPC resonator, LE IDC resonator, and CPW resonator are measured on three separate chips during three separate cooldowns to 100 mK in an adiabatic demagnetization refrigerator. Device details are shown in Table~\ref{tab:meas}. Fig.~\ref{fig:measAlOx} shows loss $\tan{\delta}$ as a function of number of photons $\langle n \rangle$ for these measurements. Each data point is determined by fitting an $S_{21}$ frequency sweep to the inverse $S_{21}$ resonator model.~\cite{Megrant2012} More details can be found in the supplementary material. Fits to the total loss model in Eqn.~\ref{eqn:totalloss} are shown as solid lines. 

From these measurement fits, we obtain the loss values in Table \ref{tab:meas}. Using Eqns.~\ref{eqn:PPCloss}, \ref{eqn:IDCloss} and \ref{eqn:CPWloss}, we obtain an inductor loss of $F_{L}\tan{\delta_L}$ = $1.12 \times 10^{-5}$ as well as a loss for the $\mathrm{Al_2O_3}$ PPC of $1.00 \times 10^{-3}$. This loss includes both the interface loss of the $\mathrm{Al/Al_2O_3/Al}$ interfaces as well as the bulk sputtered $\mathrm{Al_2O_3}$ loss. Due to the high vacuum in situ growth of the trilayer, we assume that the interfaces are much less lossy than the bulk, and thus the loss is largely a representation of the sputtered $\mathrm{Al_2O_3}$ loss.

A simpler and more commonly used method of determining TLS loss of a component of interest is to measure a resonator with that component included in it, say, as the capacitor, and then assigning all measured loss to that component; i.e., the single measurement technique. We can compare the extracted PPC value above to the value from the single measurement technique, $F_A\tan{\delta_A}$ = $9.20 \times 10^{-4}$.

The fractional difference between the total loss of the resonator $\tan\delta_{\mathrm{tot}}$ and the loss of the component of interest $\tan\delta_{C}$ is the systematic error in the single measurement technique over the dielectric loss extraction method:
\begin{equation}
    \sigma_{\mathrm{err}} = (F_{\mathrm{tot}}\tan\delta_{\mathrm{tot}} - F_{C}\tan\delta_{C})/F_{\mathrm{tot}}\tan\delta_{\mathrm{tot}}.
\end{equation}
The magnitude of $\sigma_{\mathrm{err}}$ depends on the participation ratio of the component of interest, as well as the losses of the component and the total resonator.

The dielectric loss extraction example in this paper is performed with losses in the mid- to high-range ($10^{-5}$ to $10^{-3}$) and an inductor with a participation ratio of 0.102 and two orders of magnitude lower loss than the capacitor. In this regime the systematic error of the single measurement technique over the  dielectric loss extraction method is $\sigma_{\mathrm{err}} =$0.11.

An outline of the various error regimes is shown in Fig.~\ref{fig:error}. Fig.~\ref{fig:error}~(a) shows the effect of mismatched losses in the capacitor and inductor when $C_L/C_{\mathrm{tot}} = 0.102$, as in this paper. When the capacitor is much lossier than the inductor, $\sigma_{\mathrm{err}}$ flattens out just above 0.11. In other words, an inductor with loss $F_{L}\tan\delta_{L} \sim 10^{-5}$ can measure capacitor loss $F_{C}\tan\delta_{C} \gtrsim 10^{-5}$ with $\sigma_{\mathrm{err}} \lesssim 0.11$. However, when the inductor is lossier than the capacitor, $\sigma_{\mathrm{err}} >> 0.1$ and the single measurement technique is no longer valid. In this regime, dielectric loss extraction would need to be performed, or $C_L/C_{\mathrm{tot}}$ would need to be decreased significantly by modifying the resonator design. The effect of this design modification is shown in Fig.~\ref{fig:error}~(b). A decrease of the participation loss of the inductor to below 0.01 would need to occur in order to measure capacitor losses significantly lower than the inductor loss with an error of 10$\%$ or lower using the single measurement technique. 

The grey boxes in Fig.~\ref{fig:error} show the regime where we are able to measure low loss materials ($\tan{\delta^0_{\mathrm{TLS}}} \lesssim 10^{-6}$) with $\sigma_{err} \lesssim 0.1$ without the use of the dielectric loss extraction method. A low loss and/or low participation inductor design is required. Possible modifications to the resonator design and their effects on participation ratios are illustrated in the supplementary material, while reducing the inductor loss can be attempted through nanofabrication techniques such as surface nitridation or using higher quality liftoff films.

The comparison above includes only the systematic error in the single measurement technique which is not present in the dielectric loss extraction method. Other types of errors exist which are common to both methods, including TLS loss variation over nominally identical resonators and over time for a single resonator. See the supplementary material for more details.

In conclusion, we demonstrate a method of TLS loss extraction by measuring a lumped element resonator fabricated from a superconductor-dielectric-superconductor trilayer.  We extract the dielectric loss by comparing to coplanar waveguide resonators and lumped element resonators with interdigitated capacitors. When demonstrating this method using measurements of resonators on a sputtered $\mathrm{Al/Al_2O_3/Al}$ trilayer, the TLS loss of sputtered $\mathrm{Al_2O_3}$ is shown to be $1.00 \times 10^{-3}$. We compare this  method  to  the  commonly used single measurement technique and observe a difference of 11$\%$ in extracted loss of the trilayer. This difference increases significantly with decreasing loss in the material of interest, requiring the use of dielectric loss extraction or specialized device design for materials losses of $10^{-6}$ or lower.

Next steps include extracting interface loss and bulk dielectric loss independently in a parallel plate capacitor by measuring a series of parallel plate capacitor lumped element resonators with varying capacitor dielectric thicknesses, as well as performing design modifications to optimize the accuracy of the single measurement technique.

\section*{Supplementary Material}

See Supplementary Material for details on resonator design and fabrication, inductor simulation and analysis, and variation in resonator measurements.

\section*{Data Availability}
The data that support the findings of this study are available from the corresponding author upon reasonable request.

\begin{acknowledgments}
We wish to acknowledge the support of the NIST Quantum Initiative, the Laboratory of Physical Sciences NEQST Program, and the Office of Naval Research. REL was supported by the NIST NRC Research Postdoctoral Associateship.
\end{acknowledgments}

\bibliography{lossextraction}

\providecommand{\noopsort}[1]{}\providecommand{\singleletter}[1]{#1}%
\begin{thebibliography}{24}%
\makeatletter
\providecommand \@ifxundefined [1]{%
 \@ifx{#1\undefined}
}%
\providecommand \@ifnum [1]{%
 \ifnum #1\expandafter \@firstoftwo
 \else \expandafter \@secondoftwo
 \fi
}%
\providecommand \@ifx [1]{%
 \ifx #1\expandafter \@firstoftwo
 \else \expandafter \@secondoftwo
 \fi
}%
\providecommand \natexlab [1]{#1}%
\providecommand \enquote  [1]{``#1''}%
\providecommand \bibnamefont  [1]{#1}%
\providecommand \bibfnamefont [1]{#1}%
\providecommand \citenamefont [1]{#1}%
\providecommand \href@noop [0]{\@secondoftwo}%
\providecommand \href [0]{\begingroup \@sanitize@url \@href}%
\providecommand \@href[1]{\@@startlink{#1}\@@href}%
\providecommand \@@href[1]{\endgroup#1\@@endlink}%
\providecommand \@sanitize@url [0]{\catcode `\\12\catcode `\$12\catcode
  `\&12\catcode `\#12\catcode `\^12\catcode `\_12\catcode `\%12\relax}%
\providecommand \@@startlink[1]{}%
\providecommand \@@endlink[0]{}%
\providecommand \url  [0]{\begingroup\@sanitize@url \@url }%
\providecommand \@url [1]{\endgroup\@href {#1}{\urlprefix }}%
\providecommand \urlprefix  [0]{URL }%
\providecommand \Eprint [0]{\href }%
\providecommand \doibase [0]{http://dx.doi.org/}%
\providecommand \selectlanguage [0]{\@gobble}%
\providecommand \bibinfo  [0]{\@secondoftwo}%
\providecommand \bibfield  [0]{\@secondoftwo}%
\providecommand \translation [1]{[#1]}%
\providecommand \BibitemOpen [0]{}%
\providecommand \bibitemStop [0]{}%
\providecommand \bibitemNoStop [0]{.\EOS\space}%
\providecommand \EOS [0]{\spacefactor3000\relax}%
\providecommand \BibitemShut  [1]{\csname bibitem#1\endcsname}%
\let\auto@bib@innerbib\@empty
\bibitem [{\citenamefont {Martinis}\ \emph {et~al.}(2005)\citenamefont
  {Martinis}, \citenamefont {Cooper}, \citenamefont {McDermott}, \citenamefont
  {Steffen}, \citenamefont {Ansmann}, \citenamefont {Osborn}, \citenamefont
  {Cicak}, \citenamefont {Oh}, \citenamefont {Pappas}, \citenamefont
  {Simmonds},\ and\ \citenamefont {Yu}}]{Martinis2005}%
  \BibitemOpen
  \bibfield  {author} {\bibinfo {author} {\bibfnamefont {J.~M.}\ \bibnamefont
  {Martinis}}, \bibinfo {author} {\bibfnamefont {K.~B.}\ \bibnamefont
  {Cooper}}, \bibinfo {author} {\bibfnamefont {R.}~\bibnamefont {McDermott}},
  \bibinfo {author} {\bibfnamefont {M.}~\bibnamefont {Steffen}}, \bibinfo
  {author} {\bibfnamefont {M.}~\bibnamefont {Ansmann}}, \bibinfo {author}
  {\bibfnamefont {K.~D.}\ \bibnamefont {Osborn}}, \bibinfo {author}
  {\bibfnamefont {K.}~\bibnamefont {Cicak}}, \bibinfo {author} {\bibfnamefont
  {S.}~\bibnamefont {Oh}}, \bibinfo {author} {\bibfnamefont {D.~P.}\
  \bibnamefont {Pappas}}, \bibinfo {author} {\bibfnamefont {R.~W.}\
  \bibnamefont {Simmonds}}, \ and\ \bibinfo {author} {\bibfnamefont {C.~C.}\
  \bibnamefont {Yu}},\ }\href {\doibase 10.1103/PhysRevLett.95.210503}
  {\bibfield  {journal} {\bibinfo  {journal} {Phys. Rev. Lett.}\ }\textbf
  {\bibinfo {volume} {95}},\ \bibinfo {pages} {210503} (\bibinfo {year}
  {2005})}\BibitemShut {NoStop}%
\bibitem [{\citenamefont {Gao}(2008)}]{Gao2008}%
  \BibitemOpen
  \bibfield  {author} {\bibinfo {author} {\bibfnamefont {J.}~\bibnamefont
  {Gao}},\ }\emph {\bibinfo {title} {{The Physics of Superconducting Microwave
  Resonators}}},\ \href {\doibase 10.1088/0031-9120/1/1/306} {Ph.D. thesis}
  (\bibinfo {year} {2008})\BibitemShut {NoStop}%
\bibitem [{\citenamefont {Calusine}\ \emph {et~al.}(2018)\citenamefont
  {Calusine}, \citenamefont {Melville}, \citenamefont {Woods}, \citenamefont
  {Das}, \citenamefont {Stull}, \citenamefont {Bolkhovsky}, \citenamefont
  {Braje}, \citenamefont {Hover}, \citenamefont {Kim}, \citenamefont {Miloshi},
  \citenamefont {Rosenberg}, \citenamefont {Sevi}, \citenamefont {Yoder},
  \citenamefont {Dauler},\ and\ \citenamefont {Oliver}}]{Calusine2018}%
  \BibitemOpen
  \bibfield  {author} {\bibinfo {author} {\bibfnamefont {G.}~\bibnamefont
  {Calusine}}, \bibinfo {author} {\bibfnamefont {A.}~\bibnamefont {Melville}},
  \bibinfo {author} {\bibfnamefont {W.}~\bibnamefont {Woods}}, \bibinfo
  {author} {\bibfnamefont {R.}~\bibnamefont {Das}}, \bibinfo {author}
  {\bibfnamefont {C.}~\bibnamefont {Stull}}, \bibinfo {author} {\bibfnamefont
  {V.}~\bibnamefont {Bolkhovsky}}, \bibinfo {author} {\bibfnamefont
  {D.}~\bibnamefont {Braje}}, \bibinfo {author} {\bibfnamefont
  {D.}~\bibnamefont {Hover}}, \bibinfo {author} {\bibfnamefont {D.~K.}\
  \bibnamefont {Kim}}, \bibinfo {author} {\bibfnamefont {X.}~\bibnamefont
  {Miloshi}}, \bibinfo {author} {\bibfnamefont {D.}~\bibnamefont {Rosenberg}},
  \bibinfo {author} {\bibfnamefont {A.}~\bibnamefont {Sevi}}, \bibinfo {author}
  {\bibfnamefont {J.~L.}\ \bibnamefont {Yoder}}, \bibinfo {author}
  {\bibfnamefont {E.}~\bibnamefont {Dauler}}, \ and\ \bibinfo {author}
  {\bibfnamefont {W.~D.}\ \bibnamefont {Oliver}},\ }\href
  {http://dx.doi.org/10.1063/1.5006888} {\bibfield  {journal} {\bibinfo
  {journal} {Appl. Phys. Lett.}\ }\textbf {\bibinfo {volume} {112}},\ \bibinfo
  {pages} {062601} (\bibinfo {year} {2018})}\BibitemShut {NoStop}%
\bibitem [{\citenamefont {Sandberg}\ \emph {et~al.}(2012)\citenamefont
  {Sandberg}, \citenamefont {Vissers}, \citenamefont {Kline}, \citenamefont
  {Weides}, \citenamefont {Gao}, \citenamefont {Wisbey},\ and\ \citenamefont
  {Pappas}}]{Sandberg2012}%
  \BibitemOpen
  \bibfield  {author} {\bibinfo {author} {\bibfnamefont {M.}~\bibnamefont
  {Sandberg}}, \bibinfo {author} {\bibfnamefont {M.~R.}\ \bibnamefont
  {Vissers}}, \bibinfo {author} {\bibfnamefont {J.~S.}\ \bibnamefont {Kline}},
  \bibinfo {author} {\bibfnamefont {M.}~\bibnamefont {Weides}}, \bibinfo
  {author} {\bibfnamefont {J.}~\bibnamefont {Gao}}, \bibinfo {author}
  {\bibfnamefont {D.~S.}\ \bibnamefont {Wisbey}}, \ and\ \bibinfo {author}
  {\bibfnamefont {D.~P.}\ \bibnamefont {Pappas}},\ }\href@noop {} {\bibfield
  {journal} {\bibinfo  {journal} {Appl. Phys. Lett.}\ }\textbf {\bibinfo
  {volume} {100}},\ \bibinfo {pages} {262605} (\bibinfo {year}
  {2012})}\BibitemShut {NoStop}%
\bibitem [{\citenamefont {Quintana}\ \emph {et~al.}(2014)\citenamefont
  {Quintana}, \citenamefont {Megrant}, \citenamefont {Chen}, \citenamefont
  {Dunsworth}, \citenamefont {Chiaro}, \citenamefont {Barends}, \citenamefont
  {Campbell}, \citenamefont {Chen}, \citenamefont {Hoi}, \citenamefont
  {Jeffrey}, \citenamefont {Kelly}, \citenamefont {Mutus}, \citenamefont
  {O'Malley}, \citenamefont {Neill}, \citenamefont {Roushan}, \citenamefont
  {Sank}, \citenamefont {Vainsencher}, \citenamefont {Wenner}, \citenamefont
  {White}, \citenamefont {Cleland},\ and\ \citenamefont
  {Martinis}}]{Quintana2014}%
  \BibitemOpen
  \bibfield  {author} {\bibinfo {author} {\bibfnamefont {C.~M.}\ \bibnamefont
  {Quintana}}, \bibinfo {author} {\bibfnamefont {A.}~\bibnamefont {Megrant}},
  \bibinfo {author} {\bibfnamefont {Z.}~\bibnamefont {Chen}}, \bibinfo {author}
  {\bibfnamefont {A.}~\bibnamefont {Dunsworth}}, \bibinfo {author}
  {\bibfnamefont {B.}~\bibnamefont {Chiaro}}, \bibinfo {author} {\bibfnamefont
  {R.}~\bibnamefont {Barends}}, \bibinfo {author} {\bibfnamefont
  {B.}~\bibnamefont {Campbell}}, \bibinfo {author} {\bibfnamefont
  {Y.}~\bibnamefont {Chen}}, \bibinfo {author} {\bibfnamefont {I.~C.}\
  \bibnamefont {Hoi}}, \bibinfo {author} {\bibfnamefont {E.}~\bibnamefont
  {Jeffrey}}, \bibinfo {author} {\bibfnamefont {J.}~\bibnamefont {Kelly}},
  \bibinfo {author} {\bibfnamefont {J.~Y.}\ \bibnamefont {Mutus}}, \bibinfo
  {author} {\bibfnamefont {P.~J.}\ \bibnamefont {O'Malley}}, \bibinfo {author}
  {\bibfnamefont {C.}~\bibnamefont {Neill}}, \bibinfo {author} {\bibfnamefont
  {P.}~\bibnamefont {Roushan}}, \bibinfo {author} {\bibfnamefont
  {D.}~\bibnamefont {Sank}}, \bibinfo {author} {\bibfnamefont {A.}~\bibnamefont
  {Vainsencher}}, \bibinfo {author} {\bibfnamefont {J.}~\bibnamefont {Wenner}},
  \bibinfo {author} {\bibfnamefont {T.~C.}\ \bibnamefont {White}}, \bibinfo
  {author} {\bibfnamefont {A.~N.}\ \bibnamefont {Cleland}}, \ and\ \bibinfo
  {author} {\bibfnamefont {J.~M.}\ \bibnamefont {Martinis}},\ }\href {\doibase
  10.1063/1.4893297} {\bibfield  {journal} {\bibinfo  {journal} {Appl. Phys.
  Lett.}\ }\textbf {\bibinfo {volume} {105}},\ \bibinfo {pages} {062601}
  (\bibinfo {year} {2014})}\BibitemShut {NoStop}%
\bibitem [{\citenamefont {Bruno}\ \emph {et~al.}(2015)\citenamefont {Bruno},
  \citenamefont {De~Lange}, \citenamefont {Asaad}, \citenamefont {Van
  Der~Enden}, \citenamefont {Langford},\ and\ \citenamefont
  {Dicarlo}}]{Bruno2015}%
  \BibitemOpen
  \bibfield  {author} {\bibinfo {author} {\bibfnamefont {A.}~\bibnamefont
  {Bruno}}, \bibinfo {author} {\bibfnamefont {G.}~\bibnamefont {De~Lange}},
  \bibinfo {author} {\bibfnamefont {S.}~\bibnamefont {Asaad}}, \bibinfo
  {author} {\bibfnamefont {K.~L.}\ \bibnamefont {Van Der~Enden}}, \bibinfo
  {author} {\bibfnamefont {N.~K.}\ \bibnamefont {Langford}}, \ and\ \bibinfo
  {author} {\bibfnamefont {L.}~\bibnamefont {Dicarlo}},\ }\href@noop {}
  {\bibfield  {journal} {\bibinfo  {journal} {Appl. Phys. Lett.}\ }\textbf
  {\bibinfo {volume} {106}},\ \bibinfo {pages} {182601} (\bibinfo {year}
  {2015})}\BibitemShut {NoStop}%
\bibitem [{\citenamefont {Wisbey}\ \emph {et~al.}(2010)\citenamefont {Wisbey},
  \citenamefont {Gao}, \citenamefont {Vissers}, \citenamefont {{Da Silva}},
  \citenamefont {Kline}, \citenamefont {Vale},\ and\ \citenamefont
  {Pappas}}]{Wisbey2010}%
  \BibitemOpen
  \bibfield  {author} {\bibinfo {author} {\bibfnamefont {D.~S.}\ \bibnamefont
  {Wisbey}}, \bibinfo {author} {\bibfnamefont {J.}~\bibnamefont {Gao}},
  \bibinfo {author} {\bibfnamefont {M.~R.}\ \bibnamefont {Vissers}}, \bibinfo
  {author} {\bibfnamefont {F.~C.~S.}\ \bibnamefont {{Da Silva}}}, \bibinfo
  {author} {\bibfnamefont {J.~S.}\ \bibnamefont {Kline}}, \bibinfo {author}
  {\bibfnamefont {L.}~\bibnamefont {Vale}}, \ and\ \bibinfo {author}
  {\bibfnamefont {D.~P.}\ \bibnamefont {Pappas}},\ }\href@noop {} {\bibfield
  {journal} {\bibinfo  {journal} {Jour. Appl. Phys.}\ }\textbf {\bibinfo
  {volume} {108}},\ \bibinfo {pages} {093918} (\bibinfo {year}
  {2010})}\BibitemShut {NoStop}%
\bibitem [{\citenamefont {Vissers}\ \emph {et~al.}(2010)\citenamefont
  {Vissers}, \citenamefont {Gao}, \citenamefont {Wisbey}, \citenamefont {Hite},
  \citenamefont {Tsuei}, \citenamefont {Corcoles}, \citenamefont {Steffen},\
  and\ \citenamefont {Pappas}}]{Vissers2010}%
  \BibitemOpen
  \bibfield  {author} {\bibinfo {author} {\bibfnamefont {M.~R.}\ \bibnamefont
  {Vissers}}, \bibinfo {author} {\bibfnamefont {J.}~\bibnamefont {Gao}},
  \bibinfo {author} {\bibfnamefont {D.~S.}\ \bibnamefont {Wisbey}}, \bibinfo
  {author} {\bibfnamefont {D.~A.}\ \bibnamefont {Hite}}, \bibinfo {author}
  {\bibfnamefont {C.~C.}\ \bibnamefont {Tsuei}}, \bibinfo {author}
  {\bibfnamefont {A.~D.}\ \bibnamefont {Corcoles}}, \bibinfo {author}
  {\bibfnamefont {M.}~\bibnamefont {Steffen}}, \ and\ \bibinfo {author}
  {\bibfnamefont {D.~P.}\ \bibnamefont {Pappas}},\ }\href {\doibase
  10.1063/1.3517252} {\bibfield  {journal} {\bibinfo  {journal} {Appl. Phys.
  Lett.}\ }\textbf {\bibinfo {volume} {97}},\ \bibinfo {pages} {232509}
  (\bibinfo {year} {2010})}\BibitemShut {NoStop}%
\bibitem [{\citenamefont {Chang}\ \emph {et~al.}(2013)\citenamefont {Chang},
  \citenamefont {Vissers}, \citenamefont {Corcoles}, \citenamefont {Sandberg},
  \citenamefont {Gao}, \citenamefont {Abraham}, \citenamefont {Chow},
  \citenamefont {Gambetta}, \citenamefont {Rothwell}, \citenamefont {Keefe},
  \citenamefont {Steffen},\ and\ \citenamefont {Pappas}}]{Chang2013}%
  \BibitemOpen
  \bibfield  {author} {\bibinfo {author} {\bibfnamefont {J.}~\bibnamefont
  {Chang}}, \bibinfo {author} {\bibfnamefont {M.~R.}\ \bibnamefont {Vissers}},
  \bibinfo {author} {\bibfnamefont {A.~D.}\ \bibnamefont {Corcoles}}, \bibinfo
  {author} {\bibfnamefont {M.}~\bibnamefont {Sandberg}}, \bibinfo {author}
  {\bibfnamefont {J.}~\bibnamefont {Gao}}, \bibinfo {author} {\bibfnamefont
  {D.~W.}\ \bibnamefont {Abraham}}, \bibinfo {author} {\bibfnamefont {J.~M.}\
  \bibnamefont {Chow}}, \bibinfo {author} {\bibfnamefont {J.~M.}\ \bibnamefont
  {Gambetta}}, \bibinfo {author} {\bibfnamefont {M.~B.}\ \bibnamefont
  {Rothwell}}, \bibinfo {author} {\bibfnamefont {G.~A.}\ \bibnamefont {Keefe}},
  \bibinfo {author} {\bibfnamefont {M.}~\bibnamefont {Steffen}}, \ and\
  \bibinfo {author} {\bibfnamefont {D.~P.}\ \bibnamefont {Pappas}},\ }\href
  {\doibase 10.1063/1.4813269} {\bibfield  {journal} {\bibinfo  {journal}
  {Appl. Phys. Lett.}\ }\textbf {\bibinfo {volume} {103}},\ \bibinfo {pages}
  {012602} (\bibinfo {year} {2013})}\BibitemShut {NoStop}%
\bibitem [{\citenamefont {Wang}\ \emph {et~al.}(2015)\citenamefont {Wang},
  \citenamefont {Axline}, \citenamefont {Gao}, \citenamefont {Brecht},
  \citenamefont {Chu}, \citenamefont {Frunzio}, \citenamefont {Devoret},\ and\
  \citenamefont {Schoelkopf}}]{Wang2015}%
  \BibitemOpen
  \bibfield  {author} {\bibinfo {author} {\bibfnamefont {C.}~\bibnamefont
  {Wang}}, \bibinfo {author} {\bibfnamefont {C.}~\bibnamefont {Axline}},
  \bibinfo {author} {\bibfnamefont {Y.~Y.}\ \bibnamefont {Gao}}, \bibinfo
  {author} {\bibfnamefont {T.}~\bibnamefont {Brecht}}, \bibinfo {author}
  {\bibfnamefont {Y.}~\bibnamefont {Chu}}, \bibinfo {author} {\bibfnamefont
  {L.}~\bibnamefont {Frunzio}}, \bibinfo {author} {\bibfnamefont {M.~H.}\
  \bibnamefont {Devoret}}, \ and\ \bibinfo {author} {\bibfnamefont {R.~J.}\
  \bibnamefont {Schoelkopf}},\ }\href@noop {} {\bibfield  {journal} {\bibinfo
  {journal} {Appl. Phys. Lett.}\ }\textbf {\bibinfo {volume} {107}},\ \bibinfo
  {pages} {162601} (\bibinfo {year} {2015})}\BibitemShut {NoStop}%
\bibitem [{\citenamefont {M{\"u}ller}, \citenamefont {Cole},\ and\
  \citenamefont {Lisenfeld}(2019)}]{Muller2017}%
  \BibitemOpen
  \bibfield  {author} {\bibinfo {author} {\bibfnamefont {C.}~\bibnamefont
  {M{\"u}ller}}, \bibinfo {author} {\bibfnamefont {J.~H.}\ \bibnamefont
  {Cole}}, \ and\ \bibinfo {author} {\bibfnamefont {J.}~\bibnamefont
  {Lisenfeld}},\ }\href {http://iopscience.iop.org/10.1088/1361-6633/ab3a7e}
  {\bibfield  {journal} {\bibinfo  {journal} {Rep. Prog. Phys}\ } (\bibinfo
  {year} {2019})}\BibitemShut {NoStop}%
\bibitem [{\citenamefont {Earnest}\ \emph {et~al.}(2018)\citenamefont
  {Earnest}, \citenamefont {B{\'{e}}janin}, \citenamefont {McConkey},
  \citenamefont {Peters}, \citenamefont {Korinek}, \citenamefont {Yuan},\ and\
  \citenamefont {Mariantoni}}]{Earnest2018}%
  \BibitemOpen
  \bibfield  {author} {\bibinfo {author} {\bibfnamefont {C.~T.}\ \bibnamefont
  {Earnest}}, \bibinfo {author} {\bibfnamefont {J.~H.}\ \bibnamefont
  {B{\'{e}}janin}}, \bibinfo {author} {\bibfnamefont {T.~G.}\ \bibnamefont
  {McConkey}}, \bibinfo {author} {\bibfnamefont {A.}~\bibnamefont {Peters}},
  \bibinfo {author} {\bibfnamefont {A.}~\bibnamefont {Korinek}}, \bibinfo
  {author} {\bibfnamefont {H.}~\bibnamefont {Yuan}}, \ and\ \bibinfo {author}
  {\bibfnamefont {M.}~\bibnamefont {Mariantoni}},\ }\href@noop {} {\bibfield
  {journal} {\bibinfo  {journal} {Supercond. Sci. Technol}\ }\textbf {\bibinfo
  {volume} {31}},\ \bibinfo {pages} {125013} (\bibinfo {year}
  {2018})}\BibitemShut {NoStop}%
\bibitem [{\citenamefont {Burnett}\ \emph {et~al.}(2018)\citenamefont
  {Burnett}, \citenamefont {Bengtsson}, \citenamefont {Niepce},\ and\
  \citenamefont {Bylander}}]{Burnett2018}%
  \BibitemOpen
  \bibfield  {author} {\bibinfo {author} {\bibfnamefont {J.}~\bibnamefont
  {Burnett}}, \bibinfo {author} {\bibfnamefont {A.}~\bibnamefont {Bengtsson}},
  \bibinfo {author} {\bibfnamefont {D.}~\bibnamefont {Niepce}}, \ and\ \bibinfo
  {author} {\bibfnamefont {J.}~\bibnamefont {Bylander}},\ }\href@noop {}
  {\bibfield  {journal} {\bibinfo  {journal} {IOP Conf. Series: Journal of
  Physics: Conf. Series}\ }\textbf {\bibinfo {volume} {969}},\ \bibinfo {pages}
  {012131} (\bibinfo {year} {2018})}\BibitemShut {NoStop}%
\bibitem [{\citenamefont {Richardson}\ \emph {et~al.}(2016)\citenamefont
  {Richardson}, \citenamefont {Siwak}, \citenamefont {Hackley}, \citenamefont
  {Keane}, \citenamefont {Robinson}, \citenamefont {Arey}, \citenamefont
  {Arslan},\ and\ \citenamefont {Palmer}}]{Richardson2016}%
  \BibitemOpen
  \bibfield  {author} {\bibinfo {author} {\bibfnamefont {C.~J.}\ \bibnamefont
  {Richardson}}, \bibinfo {author} {\bibfnamefont {N.~P.}\ \bibnamefont
  {Siwak}}, \bibinfo {author} {\bibfnamefont {J.}~\bibnamefont {Hackley}},
  \bibinfo {author} {\bibfnamefont {Z.~K.}\ \bibnamefont {Keane}}, \bibinfo
  {author} {\bibfnamefont {J.~E.}\ \bibnamefont {Robinson}}, \bibinfo {author}
  {\bibfnamefont {B.}~\bibnamefont {Arey}}, \bibinfo {author} {\bibfnamefont
  {I.}~\bibnamefont {Arslan}}, \ and\ \bibinfo {author} {\bibfnamefont {B.~S.}\
  \bibnamefont {Palmer}},\ }\href@noop {} {\bibfield  {journal} {\bibinfo
  {journal} {Supercond. Sci. Tech}\ }\textbf {\bibinfo {volume} {29}},\
  \bibinfo {pages} {064003} (\bibinfo {year} {2016})}\BibitemShut {NoStop}%
\bibitem [{\citenamefont {Wang}\ \emph {et~al.}(2009)\citenamefont {Wang},
  \citenamefont {Hofheinz}, \citenamefont {Wenner}, \citenamefont {Ansmann},
  \citenamefont {Bialczak}, \citenamefont {Lenander}, \citenamefont {Lucero},
  \citenamefont {Neeley}, \citenamefont {O'Connell}, \citenamefont {Sank},
  \citenamefont {Weides}, \citenamefont {Cleland},\ and\ \citenamefont
  {Martinis}}]{Wang2009}%
  \BibitemOpen
  \bibfield  {author} {\bibinfo {author} {\bibfnamefont {H.}~\bibnamefont
  {Wang}}, \bibinfo {author} {\bibfnamefont {M.}~\bibnamefont {Hofheinz}},
  \bibinfo {author} {\bibfnamefont {J.}~\bibnamefont {Wenner}}, \bibinfo
  {author} {\bibfnamefont {M.}~\bibnamefont {Ansmann}}, \bibinfo {author}
  {\bibfnamefont {R.~C.}\ \bibnamefont {Bialczak}}, \bibinfo {author}
  {\bibfnamefont {M.}~\bibnamefont {Lenander}}, \bibinfo {author}
  {\bibfnamefont {E.}~\bibnamefont {Lucero}}, \bibinfo {author} {\bibfnamefont
  {M.}~\bibnamefont {Neeley}}, \bibinfo {author} {\bibfnamefont {A.~D.}\
  \bibnamefont {O'Connell}}, \bibinfo {author} {\bibfnamefont {D.}~\bibnamefont
  {Sank}}, \bibinfo {author} {\bibfnamefont {M.}~\bibnamefont {Weides}},
  \bibinfo {author} {\bibfnamefont {A.~N.}\ \bibnamefont {Cleland}}, \ and\
  \bibinfo {author} {\bibfnamefont {J.~M.}\ \bibnamefont {Martinis}},\
  }\href@noop {} {\bibfield  {journal} {\bibinfo  {journal} {Appl. Phys.
  Lett.}\ }\textbf {\bibinfo {volume} {95}},\ \bibinfo {pages} {233508}
  (\bibinfo {year} {2009})}\BibitemShut {NoStop}%
\bibitem [{\citenamefont {Sage}\ \emph {et~al.}(2011)\citenamefont {Sage},
  \citenamefont {Bolkhovsky}, \citenamefont {Oliver}, \citenamefont {Turek},\
  and\ \citenamefont {Welander}}]{Sage2011}%
  \BibitemOpen
  \bibfield  {author} {\bibinfo {author} {\bibfnamefont {J.~M.}\ \bibnamefont
  {Sage}}, \bibinfo {author} {\bibfnamefont {V.}~\bibnamefont {Bolkhovsky}},
  \bibinfo {author} {\bibfnamefont {W.~D.}\ \bibnamefont {Oliver}}, \bibinfo
  {author} {\bibfnamefont {B.}~\bibnamefont {Turek}}, \ and\ \bibinfo {author}
  {\bibfnamefont {P.~B.}\ \bibnamefont {Welander}},\ }\href@noop {} {\bibfield
  {journal} {\bibinfo  {journal} {Jour. Appl. Phys.}\ }\textbf {\bibinfo
  {volume} {109}},\ \bibinfo {pages} {063915} (\bibinfo {year}
  {2011})}\BibitemShut {NoStop}%
\bibitem [{\citenamefont {Megrant}\ \emph {et~al.}(2012)\citenamefont
  {Megrant}, \citenamefont {Neill}, \citenamefont {Barends}, \citenamefont
  {Chiaro}, \citenamefont {Chen}, \citenamefont {Feigl}, \citenamefont {Kelly},
  \citenamefont {Lucero}, \citenamefont {Mariantoni}, \citenamefont {Malley},
  \citenamefont {Sank}, \citenamefont {Vainsencher}, \citenamefont {Wenner},
  \citenamefont {White}, \citenamefont {Yin}, \citenamefont {Zhao},
  \citenamefont {Palmstr{\o}m},\ and\ \citenamefont {Martinis}}]{Megrant2012}%
  \BibitemOpen
  \bibfield  {author} {\bibinfo {author} {\bibfnamefont {A.}~\bibnamefont
  {Megrant}}, \bibinfo {author} {\bibfnamefont {C.}~\bibnamefont {Neill}},
  \bibinfo {author} {\bibfnamefont {R.}~\bibnamefont {Barends}}, \bibinfo
  {author} {\bibfnamefont {B.}~\bibnamefont {Chiaro}}, \bibinfo {author}
  {\bibfnamefont {Y.}~\bibnamefont {Chen}}, \bibinfo {author} {\bibfnamefont
  {L.}~\bibnamefont {Feigl}}, \bibinfo {author} {\bibfnamefont
  {J.}~\bibnamefont {Kelly}}, \bibinfo {author} {\bibfnamefont
  {E.}~\bibnamefont {Lucero}}, \bibinfo {author} {\bibfnamefont
  {M.}~\bibnamefont {Mariantoni}}, \bibinfo {author} {\bibfnamefont {P.~J.
  J.~O.}\ \bibnamefont {Malley}}, \bibinfo {author} {\bibfnamefont
  {D.}~\bibnamefont {Sank}}, \bibinfo {author} {\bibfnamefont {A.}~\bibnamefont
  {Vainsencher}}, \bibinfo {author} {\bibfnamefont {J.}~\bibnamefont {Wenner}},
  \bibinfo {author} {\bibfnamefont {T.~C.}\ \bibnamefont {White}}, \bibinfo
  {author} {\bibfnamefont {Y.}~\bibnamefont {Yin}}, \bibinfo {author}
  {\bibfnamefont {J.}~\bibnamefont {Zhao}}, \bibinfo {author} {\bibfnamefont
  {C.~J.}\ \bibnamefont {Palmstr{\o}m}}, \ and\ \bibinfo {author}
  {\bibfnamefont {J.~M.}\ \bibnamefont {Martinis}},\ }\href@noop {} {\bibfield
  {journal} {\bibinfo  {journal} {Appl. Phys. Lett.}\ }\textbf {\bibinfo
  {volume} {100}},\ \bibinfo {pages} {113510} (\bibinfo {year}
  {2012})}\BibitemShut {NoStop}%
\bibitem [{\citenamefont {Vissers}\ \emph {et~al.}(2012)\citenamefont
  {Vissers}, \citenamefont {Weides}, \citenamefont {Kline}, \citenamefont
  {Sandberg},\ and\ \citenamefont {Pappas}}]{Vissers2012}%
  \BibitemOpen
  \bibfield  {author} {\bibinfo {author} {\bibfnamefont {M.~R.}\ \bibnamefont
  {Vissers}}, \bibinfo {author} {\bibfnamefont {M.~P.}\ \bibnamefont {Weides}},
  \bibinfo {author} {\bibfnamefont {J.~S.}\ \bibnamefont {Kline}}, \bibinfo
  {author} {\bibfnamefont {M.}~\bibnamefont {Sandberg}}, \ and\ \bibinfo
  {author} {\bibfnamefont {D.~P.}\ \bibnamefont {Pappas}},\ }\href@noop {}
  {\bibfield  {journal} {\bibinfo  {journal} {Appl. Phys. Lett.}\ }\textbf
  {\bibinfo {volume} {101}},\ \bibinfo {pages} {022601} (\bibinfo {year}
  {2012})}\BibitemShut {NoStop}%
\bibitem [{\citenamefont {O'Connell}\ \emph {et~al.}(2008)\citenamefont
  {O'Connell}, \citenamefont {Ansmann}, \citenamefont {Bialczak}, \citenamefont
  {Hofheinz}, \citenamefont {Katz}, \citenamefont {Lucero}, \citenamefont
  {McKenney}, \citenamefont {Neeley}, \citenamefont {Wang}, \citenamefont
  {Weig}, \citenamefont {Cleland},\ and\ \citenamefont
  {Martinis}}]{OConnell2008}%
  \BibitemOpen
  \bibfield  {author} {\bibinfo {author} {\bibfnamefont {A.~D.}\ \bibnamefont
  {O'Connell}}, \bibinfo {author} {\bibfnamefont {M.}~\bibnamefont {Ansmann}},
  \bibinfo {author} {\bibfnamefont {R.~C.}\ \bibnamefont {Bialczak}}, \bibinfo
  {author} {\bibfnamefont {M.}~\bibnamefont {Hofheinz}}, \bibinfo {author}
  {\bibfnamefont {N.}~\bibnamefont {Katz}}, \bibinfo {author} {\bibfnamefont
  {E.}~\bibnamefont {Lucero}}, \bibinfo {author} {\bibfnamefont
  {C.}~\bibnamefont {McKenney}}, \bibinfo {author} {\bibfnamefont
  {M.}~\bibnamefont {Neeley}}, \bibinfo {author} {\bibfnamefont
  {H.}~\bibnamefont {Wang}}, \bibinfo {author} {\bibfnamefont {E.~M.}\
  \bibnamefont {Weig}}, \bibinfo {author} {\bibfnamefont {a.~N.}\ \bibnamefont
  {Cleland}}, \ and\ \bibinfo {author} {\bibfnamefont {J.~M.}\ \bibnamefont
  {Martinis}},\ }\href@noop {} {\bibfield  {journal} {\bibinfo  {journal}
  {Appl. Phys. Lett.}\ }\textbf {\bibinfo {volume} {92}},\ \bibinfo {pages}
  {112903} (\bibinfo {year} {2008})}\BibitemShut {NoStop}%
\bibitem [{\citenamefont {McRae}\ \emph {et~al.}(2018)\citenamefont {McRae},
  \citenamefont {B{\'{e}}janin}, \citenamefont {Earnest}, \citenamefont
  {McConkey}, \citenamefont {Rinehart}, \citenamefont {Deimert}, \citenamefont
  {Thomas}, \citenamefont {Wasilewski},\ and\ \citenamefont
  {Mariantoni}}]{McRae2018}%
  \BibitemOpen
  \bibfield  {author} {\bibinfo {author} {\bibfnamefont {C.~R.}\ \bibnamefont
  {McRae}}, \bibinfo {author} {\bibfnamefont {J.~H.}\ \bibnamefont
  {B{\'{e}}janin}}, \bibinfo {author} {\bibfnamefont {C.~T.}\ \bibnamefont
  {Earnest}}, \bibinfo {author} {\bibfnamefont {T.~G.}\ \bibnamefont
  {McConkey}}, \bibinfo {author} {\bibfnamefont {J.~R.}\ \bibnamefont
  {Rinehart}}, \bibinfo {author} {\bibfnamefont {C.}~\bibnamefont {Deimert}},
  \bibinfo {author} {\bibfnamefont {J.~P.}\ \bibnamefont {Thomas}}, \bibinfo
  {author} {\bibfnamefont {Z.~R.}\ \bibnamefont {Wasilewski}}, \ and\ \bibinfo
  {author} {\bibfnamefont {M.}~\bibnamefont {Mariantoni}},\ }\href@noop {}
  {\bibfield  {journal} {\bibinfo  {journal} {Jour. Appl. Phys.}\ }\textbf
  {\bibinfo {volume} {123}} (\bibinfo {year} {2018})}\BibitemShut {NoStop}%
\bibitem [{\citenamefont {Cho}\ \emph {et~al.}(2013)\citenamefont {Cho},
  \citenamefont {Patel}, \citenamefont {Podkaminer}, \citenamefont {Gao},
  \citenamefont {Folkman}, \citenamefont {Bark}, \citenamefont {Lee},
  \citenamefont {Zhang}, \citenamefont {Pan}, \citenamefont {McDermott},\ and\
  \citenamefont {Eom}}]{Cho2013}%
  \BibitemOpen
  \bibfield  {author} {\bibinfo {author} {\bibfnamefont {K.~H.}\ \bibnamefont
  {Cho}}, \bibinfo {author} {\bibfnamefont {U.}~\bibnamefont {Patel}}, \bibinfo
  {author} {\bibfnamefont {J.}~\bibnamefont {Podkaminer}}, \bibinfo {author}
  {\bibfnamefont {Y.}~\bibnamefont {Gao}}, \bibinfo {author} {\bibfnamefont
  {C.~M.}\ \bibnamefont {Folkman}}, \bibinfo {author} {\bibfnamefont {C.~W.}\
  \bibnamefont {Bark}}, \bibinfo {author} {\bibfnamefont {S.}~\bibnamefont
  {Lee}}, \bibinfo {author} {\bibfnamefont {Y.}~\bibnamefont {Zhang}}, \bibinfo
  {author} {\bibfnamefont {X.~Q.}\ \bibnamefont {Pan}}, \bibinfo {author}
  {\bibfnamefont {R.}~\bibnamefont {McDermott}}, \ and\ \bibinfo {author}
  {\bibfnamefont {C.~B.}\ \bibnamefont {Eom}},\ }\href@noop {} {\bibfield
  {journal} {\bibinfo  {journal} {APL Materials}\ }\textbf {\bibinfo {volume}
  {1}},\ \bibinfo {pages} {042115} (\bibinfo {year} {2013})}\BibitemShut
  {NoStop}%
\bibitem [{\citenamefont {Deng}, \citenamefont {Otto},\ and\ \citenamefont
  {Lupascu}(2014)}]{Deng2014}%
  \BibitemOpen
  \bibfield  {author} {\bibinfo {author} {\bibfnamefont {C.}~\bibnamefont
  {Deng}}, \bibinfo {author} {\bibfnamefont {M.}~\bibnamefont {Otto}}, \ and\
  \bibinfo {author} {\bibfnamefont {A.}~\bibnamefont {Lupascu}},\ }\href@noop
  {} {\bibfield  {journal} {\bibinfo  {journal} {Appl. Phys. Lett.}\ }\textbf
  {\bibinfo {volume} {104}},\ \bibinfo {pages} {043506} (\bibinfo {year}
  {2014})}\BibitemShut {NoStop}%
\bibitem [{\citenamefont {Weber}\ \emph {et~al.}(2011)\citenamefont {Weber},
  \citenamefont {Murch}, \citenamefont {Slichter}, \citenamefont {Vijay},\ and\
  \citenamefont {Siddiqi}}]{Weber2011}%
  \BibitemOpen
  \bibfield  {author} {\bibinfo {author} {\bibfnamefont {S.~J.}\ \bibnamefont
  {Weber}}, \bibinfo {author} {\bibfnamefont {K.~W.}\ \bibnamefont {Murch}},
  \bibinfo {author} {\bibfnamefont {D.~H.}\ \bibnamefont {Slichter}}, \bibinfo
  {author} {\bibfnamefont {R.}~\bibnamefont {Vijay}}, \ and\ \bibinfo {author}
  {\bibfnamefont {I.}~\bibnamefont {Siddiqi}},\ }\href@noop {} {\bibfield
  {journal} {\bibinfo  {journal} {Appl. Phys. Lett.}\ }\textbf {\bibinfo
  {volume} {98}},\ \bibinfo {pages} {172510} (\bibinfo {year}
  {2011})}\BibitemShut {NoStop}%
\bibitem [{\citenamefont {Bahl}(2013)}]{Bahl2013}%
  \BibitemOpen
  \bibfield  {author} {\bibinfo {author} {\bibfnamefont {I.}~\bibnamefont
  {Bahl}},\ }\href {\doibase 10.1017/CBO9781107415324.004} {\emph {\bibinfo
  {title} {Artech House}}}\ (\bibinfo {year} {2013})\ pp.\ \bibinfo {pages}
  {230--232}\BibitemShut {NoStop}%
\end{thebibliography}%

\end{document}